\documentclass[aps,prb,floatfix,twocolumn,showpacs,preprintnumbers,superscriptaddress,amsmath,amssymb]{revtex4-1}
\usepackage[utf8]{inputenc}
\usepackage{graphicx}  % Include figure files
\usepackage{dcolumn}   % Align table columns on decimal point
\usepackage{bm}        % bold math
\usepackage{stmaryrd}  % for mapsfrom symbol
\usepackage{multirow}  % multirow tabular
\usepackage{color}     % color letters
\usepackage{amsmath,amssymb}
\usepackage{enumerate}
\usepackage{float}
\usepackage{subfigure} % Allow subfigures
\usepackage{soul}
\usepackage[hyperindex,breaklinks]{hyperref}
\usepackage{breakurl}

\newcommand{\av}[1]{\left\langle #1 \right\rangle}

\def  \BiSe      {{Bi$_2$Se$_3$}}    % Bi2Se3
\def  \BiTe      {{Bi$_2$Te$_3$}}    % Bi2Te3
\def  \Tc        {{T_{\rm C}}}       % Curie temperature
\def  \atper     {{\rm at}\%}        % atomic percent 
\def  \MnBi      {{Mn$_{\rm Bi}$}}   % substitutional position of Mn
\def  \Mni       {{Mn$_{\rm i}$}}    % interstitial position of Mn
\def  \DefBiSe   {{Bi$_{\rm Se}$}}   % intersite deffect Se->Bi (Bi-rich)
\def  \DefSeBi   {{Se$_{\rm Bi}$}}   % intersite deffect Bi->Se (Se-rich)
\def  \VacSe     {{V$_{\rm Se}$}}    % vacation in the Se position
\def  \EF        {{$E_{\rm F}$}}       % Fermi energy
\def  \Bxc       {{\Delta}}      % exchange correlation potential
\def  \bhS       {{\hat{\bm e}}}     % bold S with hat - Josef recommends e (spin unit vector)
\def  \br        {{\bm R}}           % bold r (position vector)
\def  \Dr        {{\Delta \br}}      % delta r 
\def  \Ham       {{\cal H}}          % caligraphic H (Hamiltonian)
\def  \xsub      {{x}}               % concentration of substitutional Mn atoms
\def  \xint      {{x_{\rm int}}}     % concentration of interatitial Mn atoms
\def  \muB       {{\mu_{\rm B}}}     % Bohr magneton
\def  \Jss        {J^{\rm ss}}         % exchange interaction between substitution Mn atoms in the same sublattice
\def  \Jds       {J^{\rm ds}}        % exchange interaction between substitution Mn atoms in different sublattices
\def  \Ji        {J^{\rm i}}         % exchange interaction between interstitial Mn atoms
\def  \Jis       {J^{\rm is}}        % exchange interaction between interstital and substitutional Mn atoms
\def  \Dop       {D}					% general magnetic dopant
\def  \sA		{\rm s_A}
\def  \sB		{\rm s_B}

\begin{document}

\preprint{APS/123-QED}

\title{Magnetic properties of Mn-doped Bi$_2$Se$_3$ topological insulators: ab initio calculations}

\author{K. Carva} \email{carva@karlov.mff.cuni.cz}
\author{P. Bal\'a\v{z}} 
\author{J. \v{S}ebesta} 
\author{I. Turek}
\affiliation{Charles University, Faculty of Mathematics and Physics, Department of Condensed Matter Physics, Ke Karlovu 5, 121 16 Prague, Czech Republic}
\author{J. Kudrnovsk\'y}
\author{F. M\'{a}ca} 
\author{V. Drchal}
\affiliation{Institute of Physics, Academy of Sciences of the Czech Republic,
Na Slovance 2, CZ-18221 Prague 8, Czech Republic}
\author{ J. Chico}
\affiliation{Peter Gr\"unberg Institut and Institute of Advanced Simulation, Forschungszentrum J\"ulich \& JARA, D-52428, J\"ulich, Germany}
\author{V. Sechovsk\'y} 
\affiliation{Charles University, Faculty of Mathematics and Physics, Department of Condensed Matter Physics, Ke Karlovu 5, 121 16 Prague, Czech Republic}
\author{J. Honolka} 
\affiliation{Institute of Physics, Academy of Sciences of the Czech Republic,
Na Slovance 2, CZ-18221 Prague 8, Czech Republic}
\date{\today}

\begin{abstract}
Doping \BiSe{} by magnetic ions represents an interesting problem since it may break the time reversal symmetry needed to maintain the topological insulator character. Mn dopants in \BiSe{} represent one of the most studied examples here. However, there is a lot of open questions regarding their magnetic ordering. 
In the experimental literature different Curie temperatures or no ferromagnetic order at all are reported for comparable Mn concentrations. This suggests that magnetic ordering phenomena are complex and highly susceptible to different growth parameters, which are known to affect material defect concentrations. 
So far theory focused on Mn dopants in one possible position, and neglected relaxation effects as well as native defects. 
We have used ab initio methods to calculate the \BiSe{} electronic structure influenced by magnetic Mn dopants, and exchange interactions between them. We have considered two possible Mn positions, the substitutional and interstitial one, and also native defects. We have found a sizable relaxation of atoms around Mn, which affects significantly magnetic interactions. Surprisingly, very strong interactions correspond to a specific position of Mn atoms separated by van der Waals gap.  Based on the calculated data we performed spin dynamics simulations to examine systematically the resulting magnetic order for various defect contents.  We have found under which conditions the experimentally measured Curie temperatures $\Tc$ can be reproduced, noticing that interstitial Mn atoms appear to be important here. Our theory predicts the change of $\Tc$ with a shift of Fermi level, which opens the way to tune the system magnetic properties by selective doping.

\end{abstract}

\pacs{}

\maketitle

\section{Introduction}
\label{Sec:Intro}

Bi-chalcogenides Bi$_{1-x}$Sb$_x$, \BiTe{}, and \BiSe{} have drawn attention of large groups of researchers not only because of their
interesting topological properties~\cite{r_10_Hasan_TI_RevMod,r_09_Zhang_TI_BiTe_BiSe_SbTe_NPhys,r_09_Hsieh_ObservDiracCone_BiTe_SbTe}, 
manifested by spin-resolved conductive surface states~\cite{r_07_Fu_Kane_3DTI}, but also due to their unique thermoelectric properties \cite{r_08_Noh_SOeff_Bi2Te3_ARPES}. 
The surface of 3D topological insulators (TIs) hosts Dirac electrons with linear dispersion relation. Due to the strong spin-orbit coupling the direction of electron movement is coupled to its spin in the vicinity of the Dirac point \cite{r_10_Hasan_TI_RevMod,r_17_Datzer_Minar_SpinStruct_UnoccBi2Se3}.
Recently, mutual interaction between these surface states and magnetism has become a subject targeted by many research studies~\cite{r_10_Hor_Hasan_FM_TI_Mn-BiTe,r_10_Chen_DiracFerm_MagDopedTI,r_13_Barde_FM_Mn-BiSe_epitaxial,r_15_Ruzicka_Holy_Mn-BiTe_properties,r_14_Vergn_Mertig_ExchInt_BiSe_BiTe_SbTe,Jian-Min_Zhang:PRB_2013,
Duming_Zhang:PRB_2012, r_15_Rakyta_Szuny_BandB_Surf_BiSe,r_16_Wolos_HS_Mn_Bi2Se3_Vac,r_12_Henk_TopChar_Dirac_Mn-BiTe,r_16_Taras_cb_MagStruct_Mn_BiSe,Carva2016_Mn-BiTe_FPTransport,Tkac2019_AnomPhCohLength_MnBiSe}.
When magnetic order is introduced into bulk TI, the time reversal symmetry is broken. 
Consequently, an energy gap at the Dirac point might be opened, which enhances electron scattering and reduces the surface conductivity. 
This mechanism might provide a unique way of surface electronic transport manipulation in future cutting-edge spintronics applications.

In an analogy to diluted magnetic semiconductors~\cite{r_14_Dietl_RMP_DilFMSC,r_10_Sato_Kud_Turek_FP_DMS}, magnetic order can be formed in the bulk of a 3D topological insulator by adding a small amount of magnetic dopants. Formation of long-range ferromagnetic order has been demonstrated experimentally in 
3D TIs \BiSe{} and \BiTe{} doped by a few percent of transition metals (Fe, Cr, Mn) with Curie temperature being in the order of few Kelvins~\cite{r_14_Vergn_Mertig_ExchInt_BiSe_BiTe_SbTe,r_16_Taras_cb_MagStruct_Mn_BiSe}.
In different experimental and theoretical works, however, the research leads to different conclusions.
The critical temperature measurements for Mn-doped bulk \BiSe{} with few \atper{} of Mn ions ranges from $\Tc \simeq 5\,{\rm K}$ up to $20\, {\rm K}$~\cite{r_16_Taras_cb_MagStruct_Mn_BiSe,r_13_Barde_FM_Mn-BiSe_epitaxial,r_16_Sanch_Springholz_Holy_NonmBandGap_Mn-BiSe}.
Moreover, in some cases the ferromagnetic state in Mn-doped \BiSe{} has not been found at all~\cite{r_08_Janic_JVej_Secho_Transp_Magn_Mn-BiSe_single,r_12_Choi_CarrierType_Mn-BiSe,r_15_Wei_TuningTranProp_Mn_Bi2Se3}, a spin glass state~\cite{r_05_Choi_Mn-doped_V_VI_SpGlass} has also been reported.
% some more examples of Tc's here: recent works
The reason of the discrepancies may stem from nontrivial 
interactions between magnetic atoms inside the bulk of 3D TIs and their environment which might host number of additional defects on different lattice positions.
%%%
% ad $T_c$ 100K
% Xu et al. thin multilayer film surface TC 100K DOI: 10.1038/NPHYS2351
% naopak Sanchez \cite{r_16Sanch_Springholz_Holy_NonmBandGap_Mn-BiSe} 10K minimalni vliv povrchoveho  magnetismu na E gap
%%%

Despite the expected opening of the TI surface state gap in the presence of ferromagnetic order, finite density of states has been observed near the Dirac point in a number of studies. This has been explained by the presence of in-gap states, based on scanning tunneling spectroscopy (STS) measurements and a phenomenological model \cite{Sessi2016_MagDopTI_gap}. The in-gap states are formed due to the hybridization of the magnetic impurity with bulk states rather than with the surface state itself \cite{Bouaziz2018_ImpurInducedStates,Carva2016_Mn-BiTe_FPTransport}. Therefore, a lot information can be learned from bulk \BiSe{}, and the bulk form is chosen to be the focus of this paper.

%% Description of the MnBiSe structure
%%%%%%%%%%%%%%%%%%%%%%%%%%%%%%%%%%%%%%%%%%%%%%%%%%%%%%%%%%%
\begin{figure}[ht!]
 \centering
 \includegraphics[width=.9\columnwidth]{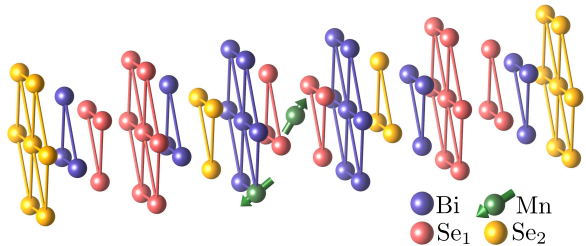}
 \caption{Hexagonal cell of the \BiSe{} structure. Possible Mn dopant positions, namely the substitutional one (\MnBi{}), and the interstitial one in the van der Waals gap (\Mni{}) are shown. }
 \label{Fig:BiSe_scheme}
\end{figure}
%%%%%%%%%%%%%%%%%%%%%%%%%%%%%%%%%%%%%%%%%%%%%%%%%%%%%%%%%%%
%%% Explain why we assumed Mn??
%%% How the surface states are modified by Mn atoms... how many percents? (Ruzicka et al)
In an ideal case, \BiSe{} has a layered structure consisting of Bi and Se hexagonal layers gathered into quintuple layers (QLs). Atomic layers inside QLs are ordered as Se(I)--Bi--Se(II)--Bi--Se(I); 
see Fig.~\ref{Fig:BiSe_scheme}. Bi layers inside QLs are equivalent, as well as the outer Se(I) layers. However,
the central Se(II) layer has apparently a different surrounding than the outer ones.
Different quintuple layers are separated by van der Waals (vdW) gaps in the crystal lattice along the $c$-axes (perpendicular to the layers) and held together by relatively weak vdW forces.
Magnetic Mn dopants can reside at different position in the crystal lattice (Fig.~\ref{Fig:BiSe_scheme}).
Mn atoms in \BiSe{} or \BiTe{} crystal lattice prefer to occupy substitutional positions (\MnBi{}) where Mn atoms replace Bi atoms    % Bi2Se3
~\cite{r_10_Hor_Hasan_FM_TI_Mn-BiTe,Jian-Min_Zhang:PRB_2013}.
In this case Mn atoms act as electron acceptors, which would induce hole-like bulk conductivity. However, experimental studies have shown an electron bulk conductivity in
\BiSe{}, which supports the assumption there must be also other doping-induced or native defects \cite{r13_Cao_bulkQHE}.
Doping-induced defects may include Mn in interstitial positions (\Mni{}), reported in several experiments~\cite{AIPAdv14_Collins,r_16_Chong_JAlCom}.
%\textcolor{blue}{(or another transition metal atoms\cite{r_12_Song_PRB86,r11_Wang_PRB84})}.
This problem has already been thoroughly studied in \BiTe{}, where interstitials appear to play a significant role \cite{r_15_Ruzicka_Holy_Mn-BiTe_properties,Carva2016_Mn-BiTe_FPTransport}. % add more references,theoretical paper    !!!
Interstitial sites in the Van der Waals gap between the quintuple layers represent the most favorable position for \Mni{} interstitials \cite{r_12_MagDopTail_BiSe}. In contrast to the substitutional Mn defects, interstitial Mn atoms are electron donors.
For \BiTe{} our previous study shows that the presence of \MnBi{} atoms shifts the Fermi level into the Mn-impurity peak in the majority spin channel located inside the gap,
while \Mni{} atoms shift the Fermi level into the conduction band~\cite{Carva2016_Mn-BiTe_FPTransport}.
Thus Mn atoms in different positions modify the bulk conductivity and, as we shall show, also exchange interactions, in significantly different ways.

Another modification of the ideal \BiSe{} structure can be caused by native defects \cite{Xia2009_NPhys_ARPES_BiSe,r13_Ruleova_Sr_doped_BiSe,r12_Shuang_defects_BiTeSe}, which occur regardless of the Mn-doping, and their formation depends on the growth conditions \cite{r03_Zunger_doping_principles,r98_Zhang_CuInSe2_defects}.
Experimentally studied samples were usually n-type, even in the presence of \MnBi{} acceptors. Samples with p-type conductivity have also been found, namely in Ca-counterdoped samples \cite{r_09_Hor_p-type_Bi2Se3_Ca} and in some of the Mn-doped samples \cite{r_12_Choi_CarrierType_Mn-BiSe,r_15_Wei_TuningTranProp_Mn_Bi2Se3, r_16_Valis_n-p_Mn-BiSe_Hetero}.
There is a general agreement that the most probable defect responsible for increasing the Fermi level are \VacSe, Se vacancies \cite{r_09_Hor_p-type_Bi2Se3_Ca,r_16_Dai_IntrinLimit_Bi2Se3,r_16_Wolos_HS_Mn_Bi2Se3_Vac}. This is supported by ab initio calculations of defect formation energies \cite{r12_West_SOI_and_defects,r_12_Scanlon_TIBulkCon_Antisite}.  
According to calculations based on the tight-binding model, vacancies are also expected to affect the TI surface state in the vicinity of the Dirac point, despite being deep below the surface ~\cite{r12_Black-Schaffer_lattmod}.
Other typical native defects in \BiSe{} are antisites, substitutions of Se-atom in the lattice by Bi (\DefBiSe{}) or {\em vice versa} (\DefSeBi{}), which lead to Bi-rich and Se-rich \BiSe{} samples, respectively.
In the former case, the excess Bi atoms act as electron acceptors. In the latter condition, where Se atoms dominate the sample, \DefSeBi{} act as electron donors.

Importantly, all the defects mentioned above influence the bulk conductivity in the system and thus modify the indirect exchange interactions between the magnetic moments of the Mn-dopants.
The authors of previous theoretical studies limited their research to magnetism in binary chalcogenides due to magnetic dopants on substitutional 
positions~\cite{r_14_Vergn_Mertig_ExchInt_BiSe_BiTe_SbTe,r_12_Henk_TopChar_Dirac_Mn-BiTe}
without considering possible coexisting interstitial magnetic moments or native defects. 
The goal of this paper is to provide a thorough picture of complex interplay between magnetic dopants in \BiSe{} on different positions and
native defect in the structure. This effort aims towards answering a question how the magnetic ground state in Mn-doped \BiSe{} is influenced
by the mentioned factors and how to reach the long-range ferromagnetic ordering of Mn magnetic moments at higher temperatures. We should note that most of the experimental analysis came to conclusion that in Mn-doped chalcogenides one does not usually observe any 
clustering of Mn atoms~\cite{r_15_Ruzicka_Holy_Mn-BiTe_properties,r_16_Taras_cb_MagStruct_Mn_BiSe}, therefore random dopant distribution can be assumed.

%%% What shall we do
Because of small amounts of magnetic atoms in the samples, the dominant exchange mechanisms between the magnetic moments are indirect RKKY interaction~\cite{r_09_Liu_MagImp_Surf_TI,r10_Wray_NatPhys,r_12_Check_Dirac_FM_Mn_BiTeSe,r16_Zimmermann_PRB94} and super-exchange~\cite{r_10_Sato_Kud_Turek_FP_DMS,r05_Dederich_DilMagSem,Ruesmann2018_ControlExchange_TISurf}. %{ \cite{r10_Yu_Science,r16_Zimmermann_PRB94}}.
Especially the first one is strongly influenced by the number of carriers in the system. Thus it depends on the ratio of substitutional and interstitial Mn atoms as well as on the concentration of antisite defects and vacancies.
By means of {\em ab initio} methods we calculated exchange interactions between magnetic moments of Mn atoms. 
In particular we studied how the interaction changes due to increasing number of \Mni{} dopants. The influence of native defects has also been taken into account. 
Finally, we have used the {\em ab initio} results as an input for atomistic Monte Carlo (MC) simulations to calculate Curie temperatures as a function of Mn concentration.

%%% Organization of the paper
The paper is organized as follows. In Section~\ref{Sec:Methods} we describe the {\em ab initio} and MC calculations.
In Sec.~\ref{Sec:Results} we introduce and analyze our results. Finally, we sum up the results and draw conclusions of our research in Sec.~\ref{Sec:Conclusions}

\section{Formalism and computational details}

% the last sentence is from our previous paper. Does it agree also for this work?

\label{Sec:Methods}

% some introduction about the methods

\subsection{{\em Ab initio} calculations}
\label{SSec:AbIniti}

Our density functional theory (DFT) calculations were based on Green function tight-binding linear muffin-tin orbital method~\cite{r_84_Skriver_LMTO_book} making use of atomic--sphere approximation (TB-LMTO-ASA)~\cite{r_97_tdk}.
To treat disorder in solid crystals we have employed the coherent potential approximation (CPA)~\cite{r_68_vke}.
This method is significantly more efficient than the supercell approach in the case of
low concentrations of defects, especially when there are several different defects with small concentration, which is the situation in the studied system.

The local spin-density approximation (LSDA) employing the Vosko-Wilk-Nusair exchange-correlation potential~\cite{r_80_vwn}, and  the $s$, $p$, $d$ basis are used. We should note that for a different material with vdW gap, Fe$_{3}$GeTe$_2$, LSDA has provided a more accurate value of magnetic moment than GGA or vdW based functionals \cite{Zhuang2016_FeGeTe_MAE_Calc}.
Bi and Se atoms are heavy elements featuring significant spin-orbit coupling. The importance of spin-orbit interaction in connection to defect states has also been demonstrated \cite{r12_West_SOI_and_defects}. Therefore, 
first order relativistic correction were included by adding the on-site spin-orbit coupling term to the scalar-relativistic
TB-LMTO Hamiltonian. The value of the spin-orbit coupling was determined self-consistently during the calculations~\cite{r_08_tdk_LMTO_SO}.
To enhance the treatment of alloy electrostatics in the case of disorder, we have also used a simple screened impurity model \cite{r_95_Korzh_Ruban_ScrImp_Madelung_CPA}.

We have used experimental lattice parameters, $a=4.138~\mathrm{\AA}$ and $c=28.64~\mathrm{\AA}$ \cite{r_10_Zhang_FPstudy_BiTe_BiSe_SbTe_NJOP}. Reciprocal space has been discretized to an uniform mesh of about $3.10^4$ points in the Brillouin zone for energies near the Fermi level, while the mesh density has been gradually reduced for states at lower energy. The formula unit consists of 2 Bi and 3 Se atoms, and we have used the rhombohedral unit cell which also contains only 5 atoms, contrary to the hexagonal cell choice. In addition to that, an empty sphere is placed at the vdW positions in \BiSe{} crystals in order to improve the space filling for the LMTO-ASA method. It is convenient to consider the vdW positions as another (sixth) sublattice, since it can host Mn interstitials. This geometry has already been successfully applied to \BiTe{}  \cite{Carva2016_Mn-BiTe_FPTransport}. 
Mn dopants are thus allowed here to form two magnetic sublattices of substitutional Mn atoms, and one magnetic sublattice of interstitial Mn atoms. \MnBi{} atoms randomly occupy two equivalent Bi sublattices with the same probability, forming {Bi$_{2-\xsub}$Mn$_\xsub$Se$_3$} alloys.
\Mni{} atoms can randomly replace the empty sphere in the vdW gap, which chemically corresponds to {Bi$_2$Se$_3$Mn$_\xint$}. 

We have included possible relaxations of the nearest atoms around \MnBi{}, whose significant impact is described in the Results section. For the relaxation of a defect neighborhood we use spin polarized DFT based on projector augmented wave (PAW) potentials as implemented in the Vienna ab initio Simulation Package (VASP) \cite{r_96_Kress_AbInitPP_PW,r_99_Kresse_PAW}. The generalized gradient approximation (GGA) of Perdew, Burke, and Ernzerhof \cite{r_96_Perdew_GGAsimple} is employed for the exchange-correlation functional. We have found only very small differences in electronic structure when compared to the LSDA based calculation. All defect calculations are performed on a hexagonal  60-atom cell. We have only one defect in the unit cell, i.e. the defect-defect distance equals the dimension of the supercell, and concentration is 1/24, approx. 4\%. This is within the expected defect concentration range, which allows for establishing the correct Fermi level position in the doped system, the most crucial effect for dopant behavior \cite{r98_Zhang_CuInSe2_defects}.  To avoid the basis set incompleteness, a large enough cutoff energy of 500 eV is used for the plane-wave expansion. The conjugated gradient technique is employed for the geometry optimization of the configurations. The geometries were relaxed with a residual force criterion of 2.5 meV/\AA . For the self-consistency cycle,we have employed an energy criterion of 10$^{–4}$ eV. A k-point mesh of 10x10x2 is used to sample the Brillouin zone. 
The distortion of the structure caused by Mn dopants could be reflected in the TB-LMTO-CPA approach by a modification of Wigner-Seitz radii \cite{r_84_Skriver_LMTO_book,r_90_kd_LMTO_CPA_diffWS,r_09_kmt_Redin_AFM_Fe_Ir001}, namely global radii $w^{all}$ related to the volume of the unit cell, and local radii $w^{Q}$ associated to the atomic specie $Q$. 
The conventional choice is $w^{Q}$=$w^{all}$ for all species occupying one site.
We have modified this choice locally for $Q$=Mn, Bi in such a way that the volume of the Bi-sublattice, 
$(1-x) (w^{\rm Bi})^{3} + x (w^{\rm Mn})^{3}$, is preserved, while $w^{\rm Mn}$ reflects the relaxed atomic surrounding obtained from VASP calculations.
According to the transformation properties of the LMTO structure constants \cite{r_84_Skriver_LMTO_book,r_90_kd_LMTO_CPA_diffWS} this leads to a corresponding modification of hopping integrals, and during the selfconsistent loop also to the change of potential parameters. The use of CPA without disorder-induced local relaxations has also been found to introduce significant errors in calculations of other materials, for example TiAlN alloys. In this system the effect of relaxation was succesfully incorporated by a different approach, the independent sublattice model, leading again to an agreement with supercell calculations \cite{Alling2007_TiAlN_Disord_Relax}. 
 Small total volume changes due to Mn-doping and native defects were neglected. 
  \DefBiSe{} and \DefSeBi{} antisites as well as \VacSe{} vacancies were also treated within CPA, with the same concentrations on each sublattice.

In order to map the problem of magnetic ordering into the classical effective Heisenberg Hamiltonian as described below in Sec. \ref{SSec:AtSim} we calculate exchange interactions between magnetic dopants $\Dop,\Dop'$ occupying sites $\br$ and $\br'$ from first principles employing the Liechtenstein formula~\cite{r_87_licht_LSDAtoExchange}. Note that because of the vertex cancellation theorem \cite{r_96_bkd} the formula can be constructed for disordered systems in an analogous way to the ordered system by considering conditionally averaged Green functions \cite{r_06_tkdb_ExchInt_CurTemp_philmag} within the CPA method:

\begin{equation}
  \begin{split}
    J^{\Dop\Dop'}_{\br\br'} = & \frac{1}{4\pi}\,\Im \int_C{\rm d}z Tr  \\
          & \left[ \Bxc^{\Dop}_\br\left(z\right) {\bar{G}}^{\Dop\Dop',\uparrow}_{\br\br'}\left(z\right)\, \Bxc^{\Dop'}_\br\left(z\right) {\bar{G}}^{\Dop\Dop',\downarrow}_{\br\br'}\left(z\right)\right]\,.
  \end{split}
\label{Eq:Liecht}
\end{equation}
Here the energy integration is performed in the upper half of the complex energy plane along a contour $C$ starting below the bottom of the valence band and ending at the
Fermi energy, the trace runs over angular momenta $L=(\ell,m)$,  
$\Bxc^{\Dop}_\br$ are matrices corresponding to the spin splitting of LMTO potential functions on dopants $\Dop$ at sites $\br$ (see \cite{r_06_tkdb_ExchInt_CurTemp_philmag,r_04_ktd_exch_DMS} for a more accurate definition). 
${\bar{G}}^{\Dop\Dop',\sigma}_{\br\br'}$ are site off-diagonal blocks of the conditionally averaged Green function, the average of the spin-dependent Green function  with spin $\sigma \in \{ \uparrow, \downarrow\}$ over all
configurations with atoms of the types $\Dop,\Dop'$ fixed at sites $\br$ and $\br'$, respectively.

Due to the translational invariance of the effective system its exchange interactions $J^{\Dop\Dop'}_{\br\br'}$ are fully described by $J^{\Dop\Dop'}(\br-\br')=J^{\Dop\Dop'}(\Dr)$. Note that within basic assumptions of CPA these quantities are independent of particular configuration, and can thus be calculated only once for each defect concentration. They of course depend on defect concentration, especially in systems where a small concentration change affects significantly states at the Fermi level.

\subsection{Atomistic simulations}
\label{SSec:AtSim}

In the second stage we employed atomistic simulations~\cite{Evans:J_Phys_2014,Hinzke2015,Skubic:J_Phys_2008,Eriksson_book2017} to calculate the Curie temperatures of given samples.
This method allows us to study basic thermodynamic properties of a material on the scale of distances between the atoms.
At this point, we define a 3-dimensional supercell of Mn atoms of size $L \times L \times L$,
where $L$ is number of elementary unit cells repeating along one direction.
In the supercell, we randomly generated substitutional \MnBi{} atoms with equal concentration, $\xsub$, on both Bi sublattices,
and \Mni{} atoms with concentration $\xint$. In our notation each magnetic dopant indexed by $i$ is thus of type $\Dop_i\in \{\sA, \sB, {\rm i}\}$, denoting the substitutional \MnBi{} on the A or B sublattice, or \Mni{} interstitials in the VdW gap. It is assigned a position $\br_i$  from the set of its parent sublattice points within the supercell.  Since the sublattices are equivalent, it is  sufficient to define $\Jss(\br)=J^{\sA \sA}(\br)=J^{\sB \sB}(\br)$ and $\Jds(\br)=J^{\sA \sB}(\br)=J^{\sB \sA}(\br)$. 
Each particular configuration with $N$ magnetic dopants is thus described by a set of $\Dop_i$ and $\br_i$ for $i=1..N$. The total exchange energy of the supercell is then given by Hamiltonian of the Heisenberg type
\begin{equation}
  \Ham_{\rm xc} = -\sum_{i,j} J^{\Dop_i\Dop_j}(\br_i-\br_j)\; \bhS_i(\br_i) \cdot \bhS_j(\br_j)\,,
\label{Eq:Hxc}
\end{equation}
where $\bhS_i(\br_i)$ and $\bhS_j(\br_j)$ are unit vectors with the directions of $i$-th and $j$-th magnetic moments, respectively, at sites occupied by Mn dopants (in the calculated particular configuration).
Note, in the atomistic simulations we did not consider any other atoms than Mn atoms, their effects is included via their influence on the exchange constants $J_{ij}$.

To estimate the critical temperature we used the Uppsala Atomistic Spin Dynamics (UppASD) code~\cite{Skubic:J_Phys_2008,r_08_Hells_AtSpinDyn_DMS,UppASD}.
We made use of the classical MC method implemented in the UppASD package utilizing Metropolis algorithm. 
The magnetic ground state was obtained by making use of the simulated annealing technique, in which the simulation is started at temperature much higher than the transition temperature and then is slowly cooled down towards $T=0$. At each temperature step we performed
from $10^5$ up to $5 \times 10^5$ MC steps. 
For each examined dopant concentration associated to a unique set of exchange interactions we generated $10$ different random Mn dopant configurations according to the given concentrations. To improve the statistics we performed in parallel $5$ simulations for each such configuration, and averaged the contributions of all of them. 

At each temperature step we evaluated heat capacity, $C$, magnetic susceptibility, $\chi$, and
Binder cumulant, $U$, defined as
\begin{equation}
  U \equiv U(L,T) = 1 - \frac{\av{M^4}}{3\, \av{M^2}^2}\,,
\label{Eq:Binder}
\end{equation}
where $M$ is total magnetization of the supercell, and $\av{\dots}$ stands for averaging over MC steps at a constant temperature.
The Binder cumulant (\ref{Eq:Binder}) allows us to reduce the finite size effect in the estimation of Curie temperature in ferromagnetic systems, which are prevalent in other methods, such as via the estimation of susceptibility and/or specific heat peaks.
The basic properties of Binder cumulant~\cite{Binder:ZPCM_1981} are: (i) When $T \to 0$, $U(L,T) \to 2/3$,
(ii) when $T \to \infty$, we obtain $U(L,T) \to 1/L$, and (iii) when $T \to \Tc$, the Binder cumulant $U(L,T) \to U^{*}$, which is an invariant with respect to the 
supercell size, $L$. Thus curves $U(L,T)$ calculated for different sizes $L$ intersect in the same point, where $T=\Tc$.

\section{Results and discussion} 
\label{Sec:Results}

%%% How did we estimate positions of atoms? relaxation of the lattice?

\subsection{Electronic structure}

%%%%%%%%%%%%%%%%%%%%%%%%%%%%%%%%%%%%%%%%%%%%%%%%%%%%%%%%%%%
%%%%%              Plots of DOS                       %%%%%    
\begin{figure*}

 \centering
 \subfigure{ \includegraphics[width=0.97\columnwidth]{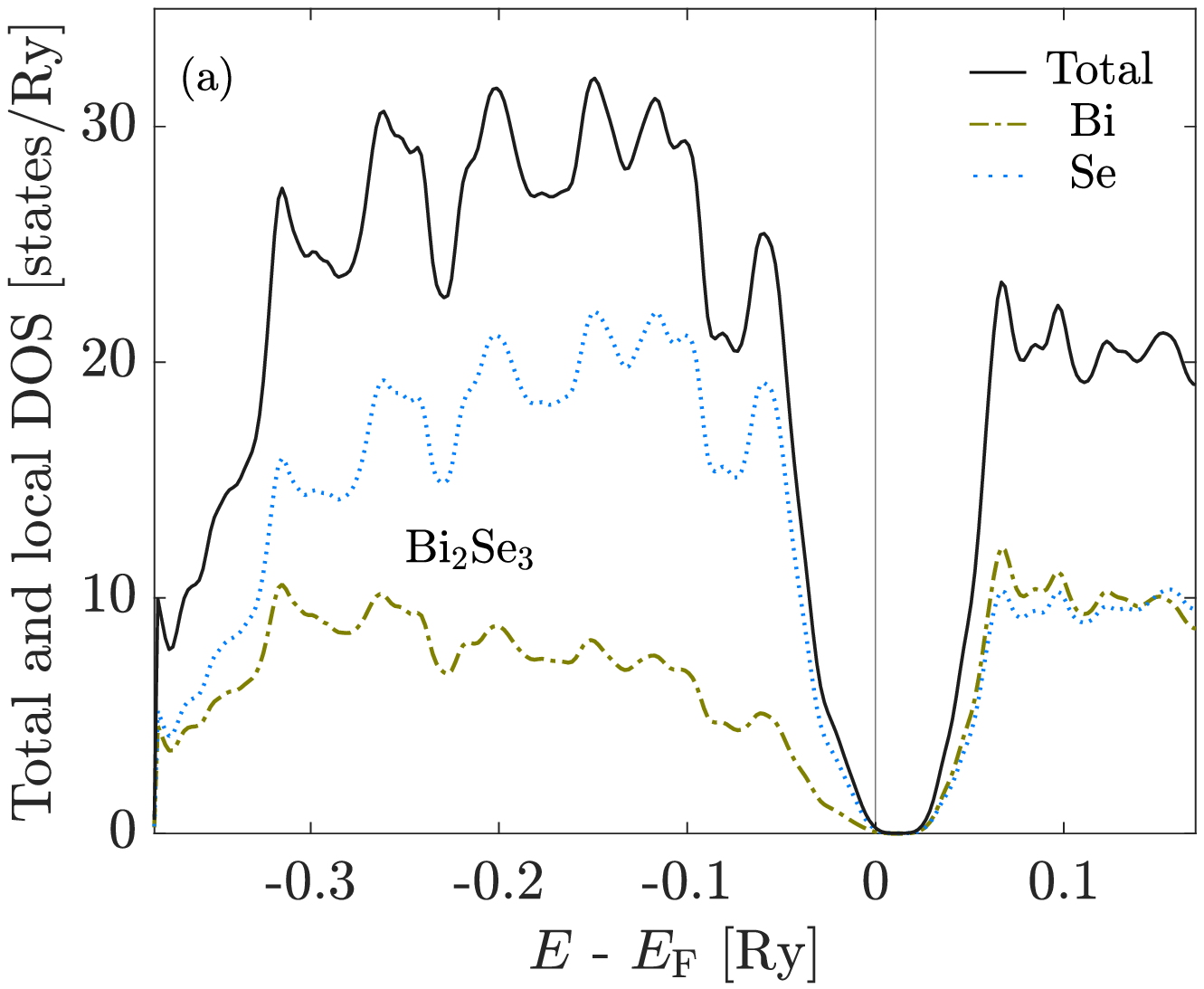}}
 \hspace{20pt}
 \subfigure{\includegraphics[width=0.97\columnwidth]{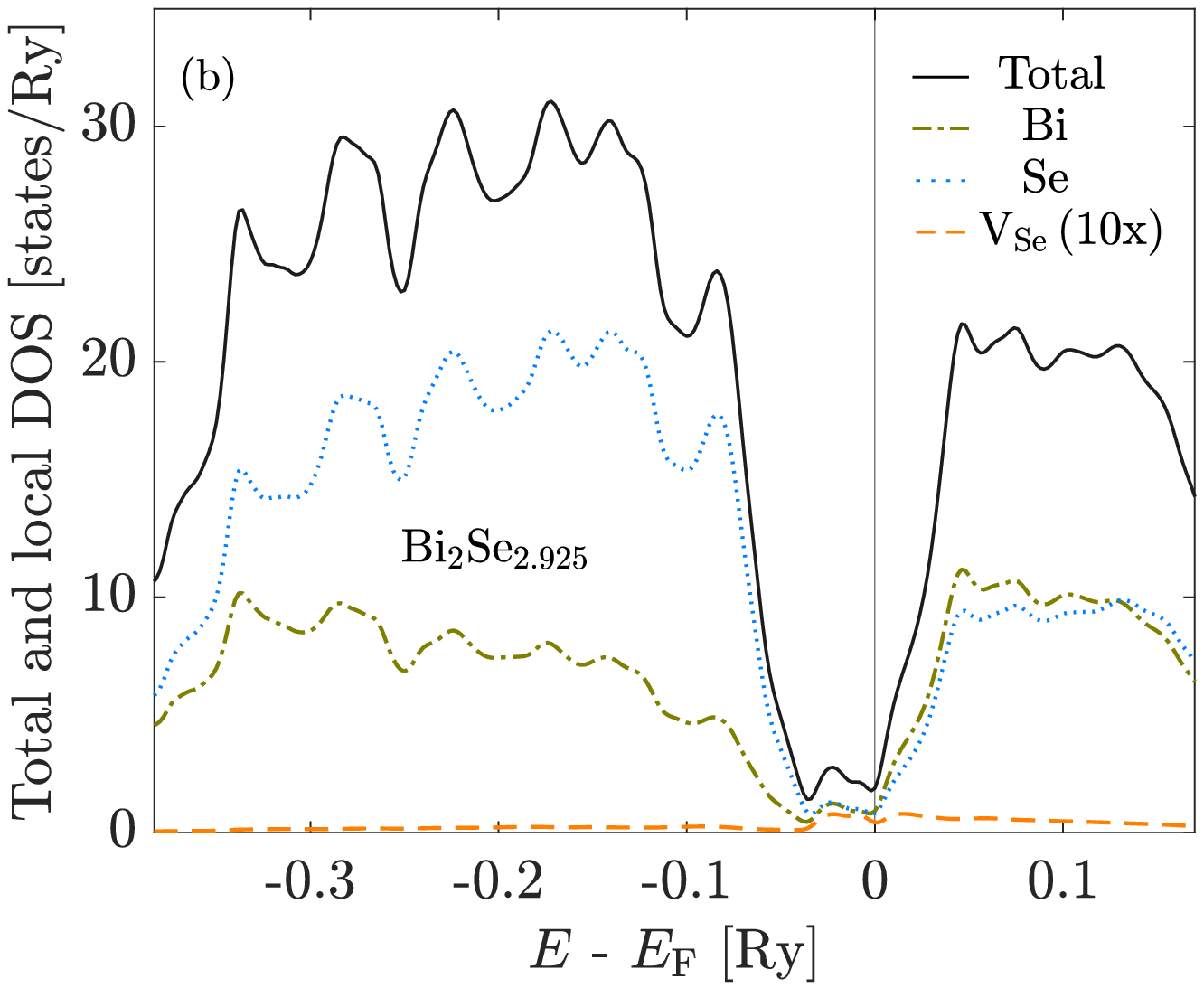}}
 
  \caption{(Color online) (a) Spin-resolved densities of states of \BiSe{}. (b) Spin resolved DOS of  \BiSe{} in the presence of vacancies on Se sites with the total concentration 0.075 per f.u.  }
  \label{Fig:DOS_pure}
\end{figure*}

\begin{figure*}
 \centering
 
 \subfigure{ \includegraphics[width=0.97\columnwidth]{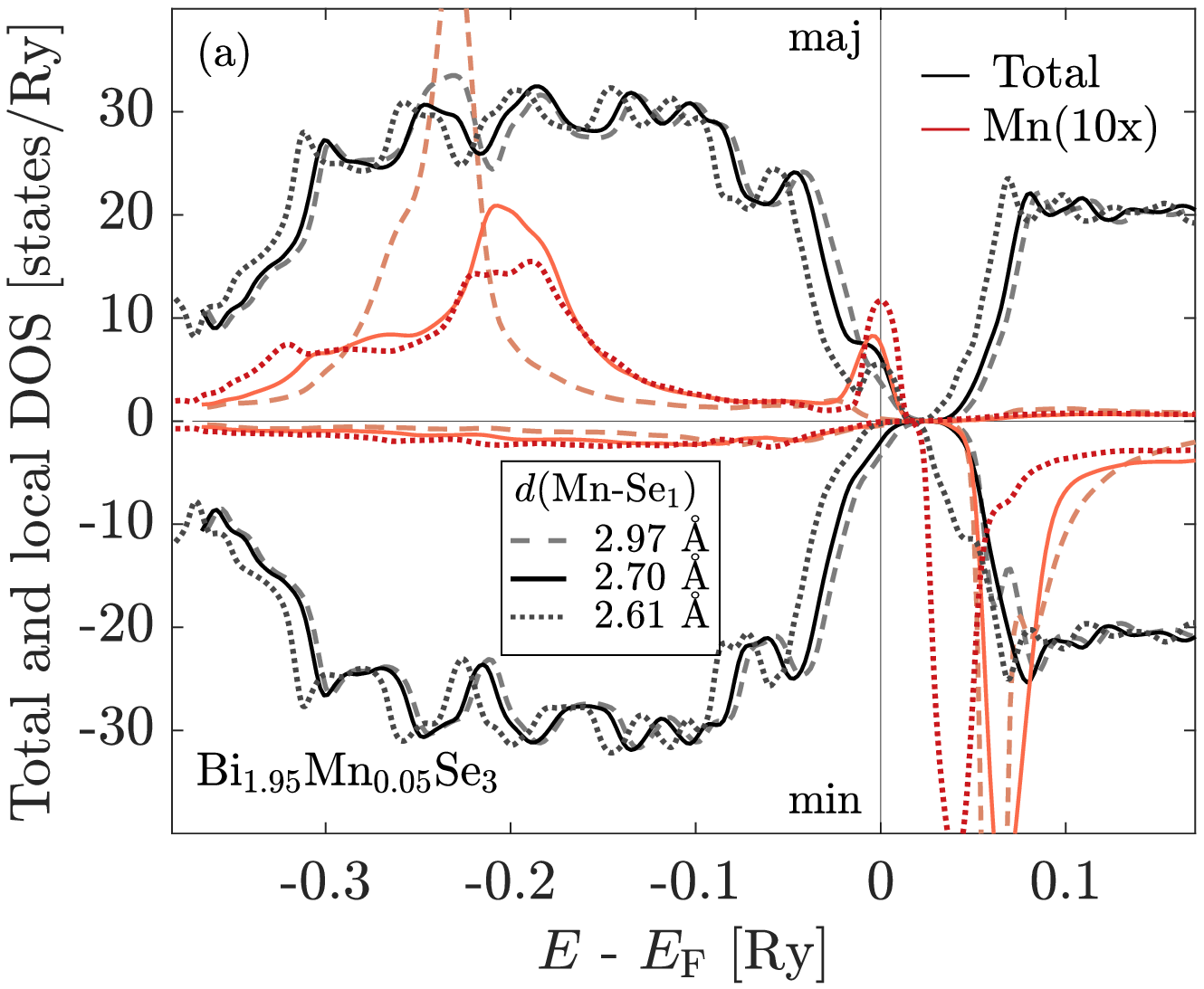}}
 \hspace{20pt}
 \subfigure{\includegraphics[width=0.97\columnwidth]{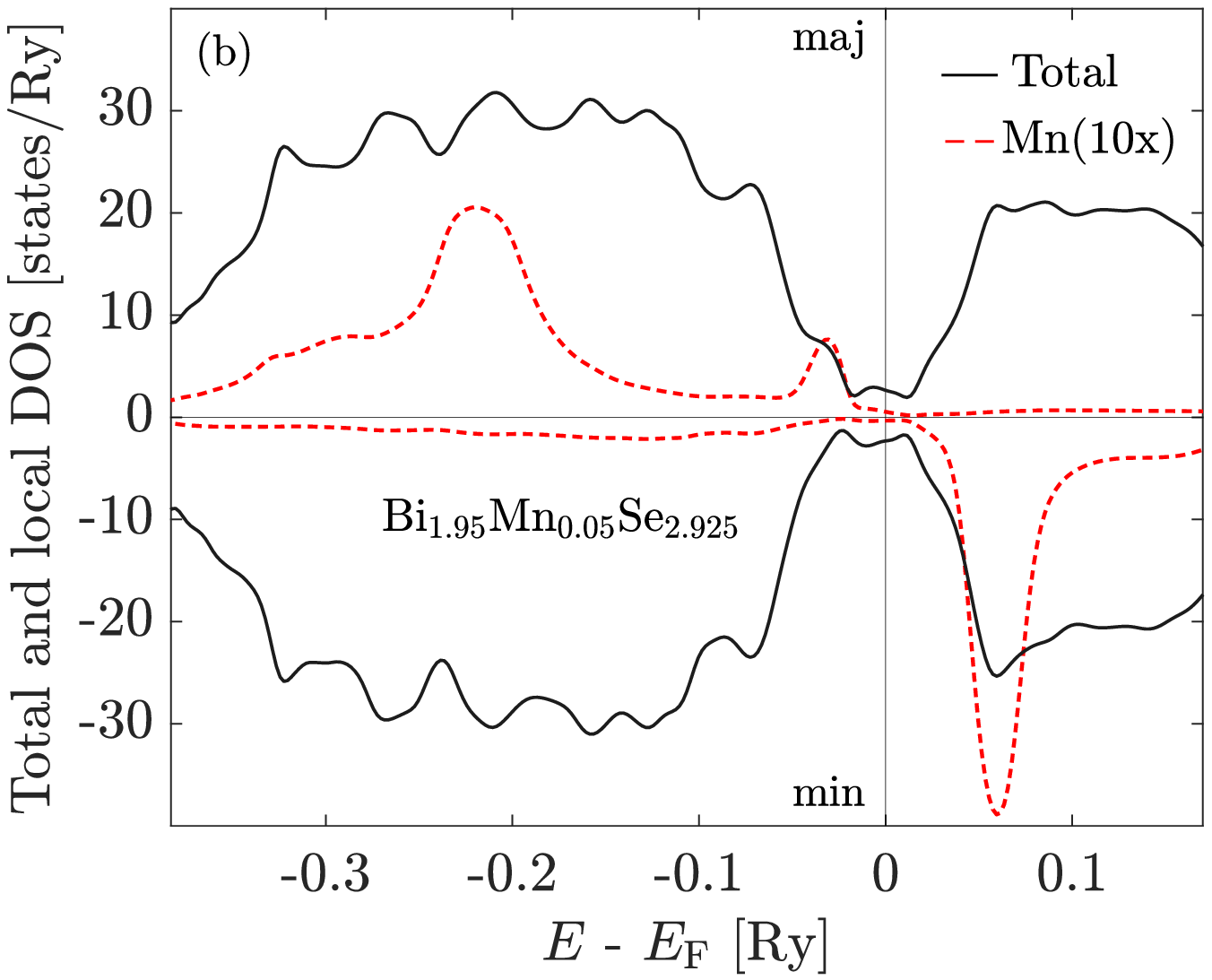}}
 
  \caption{(Color online) (a) Spin-resolved densities of states of \BiSe{} doped with \MnBi{} acceptor with the concentration $x$=0.05 per f.u. Solutions with three different local Mn radii $w^{\mathrm{Mn}}$ are depicted, and denoted in terms of the effective distance between Mn and the nearest Se. (b) Spin-resolved densities of states of \BiSe{} doped with \MnBi{} acceptor with the concentration $x$=0.05 per f.u. modified additionally by the presence of vacancies on Se sites with the total concentration 0.075 per f.u. }
  \label{Fig:DOS_sub}
\end{figure*}

\begin{figure*}
 \centering
 \subfigure{ \includegraphics[width=0.97\columnwidth]{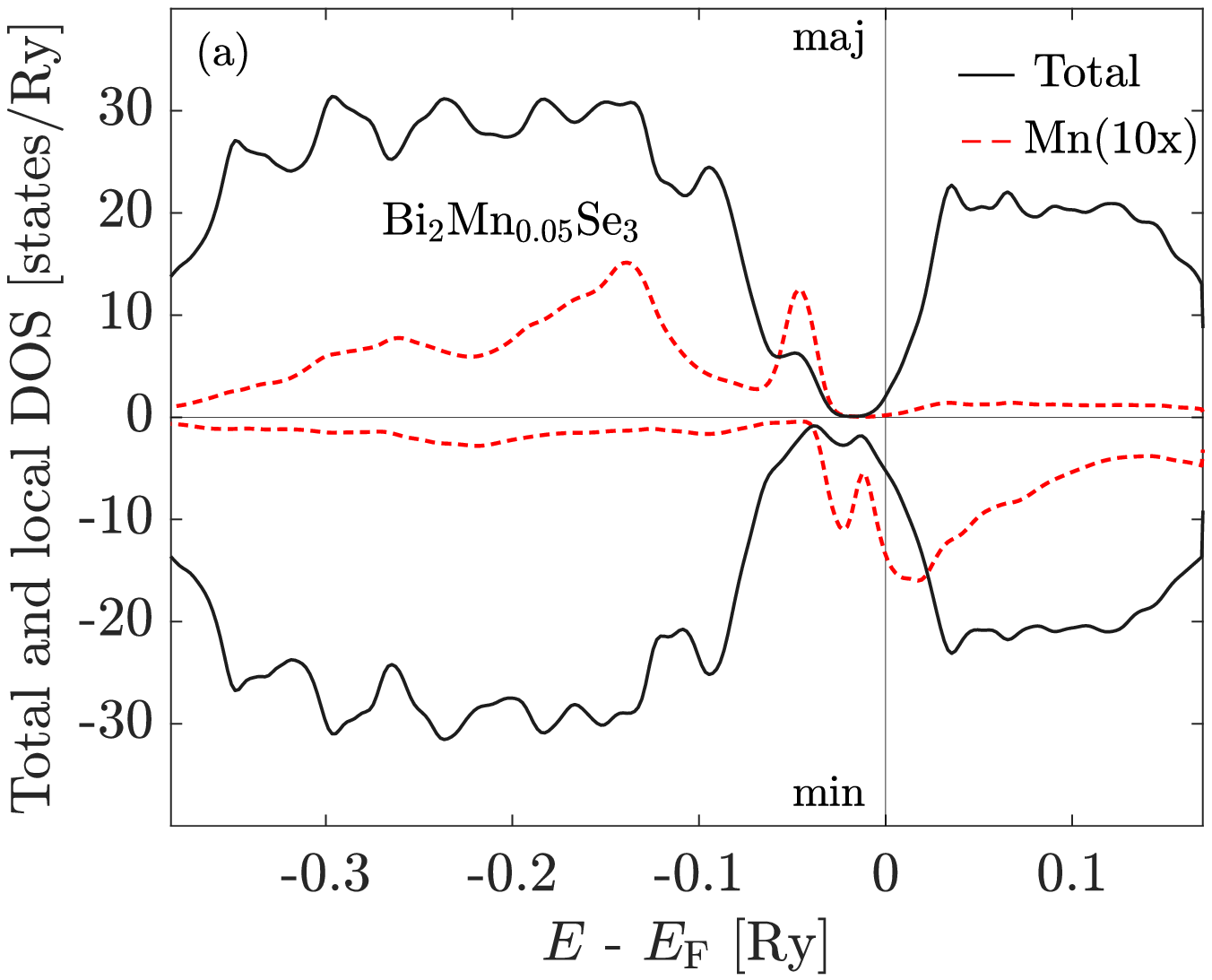}}
 \hspace{20pt}
 \subfigure{\includegraphics[width=0.97\columnwidth]{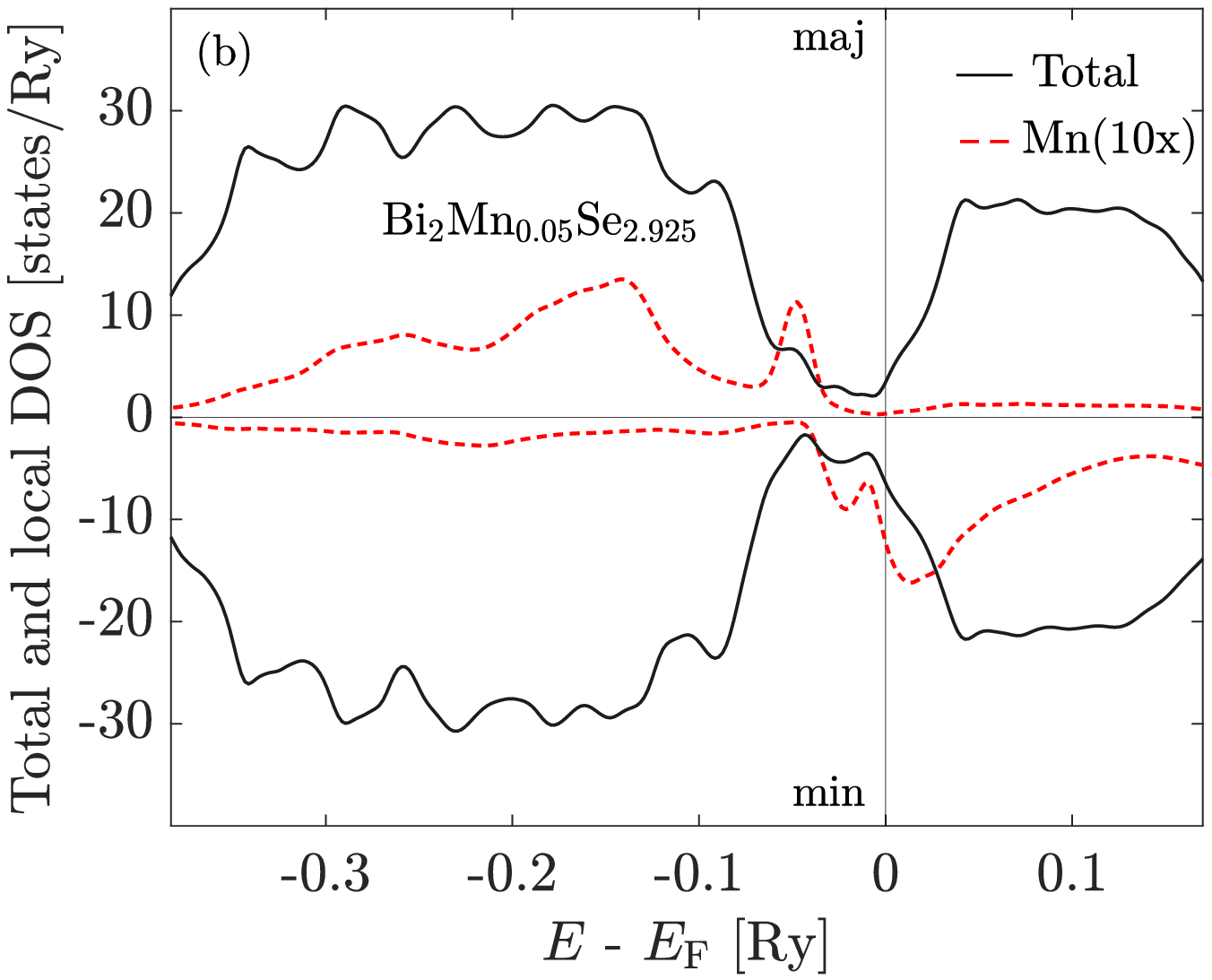}}
 
  \caption{(Color online) (a) Spin-resolved densities of states of \BiSe{} doped with \Mni{} donors with the concentration $x$=0.05. (b) The same for the system in (a) modified additionally by the presence of vacancies on Se sites with the total concentration 0.075 per f.u. }
  \label{Fig:DOS_int}
\end{figure*}

%%%%%%%%%%%%%%%%%%%%%%%%%%%%%%%%%%%%%%%%%%%%%%%%%%%%%%%%%%
%%%% Effect of the relaxation Mn sites on the DOS_int
% \begin{figure}[ht!]

%%%%%%%%%%%%%%%%%%%%%%%%%%%%%%%%%%%%%%%%%%%%%%%%%%%%%%%%%%%

Here we present the calculated densities of states (DOS's)
for systems with the considered Mn dopants of two different characters, \MnBi{} and \Mni{} (assuming always the same concentration of Mn per formula unit i.e. $x=0.05$), \VacSe{} defects, and their combinations.

Let us start with the results for the pure \BiSe{} compound without any native defect. The DOS (Fig.~\ref{Fig:DOS_pure} (a)) shows that it is an insulator with the band gap of 0.39~eV, a large value compared to other bismuth-chalcogenides~\cite{r_10_Hasan_TI_RevMod,r_09_Zhang_TI_BiTe_BiSe_SbTe_NPhys}. The obtained gap size is in good agreement with the experimentally determined one (0.33~eV)\cite{PRB13_Nechaev}.
%PHYSICAL REVIEW B 87, 121111(R) (2013)
In the valence band the Se-4$p$ electrons prevail, whereas Bi-6$p$ states dominate in the conduction band, as has already been shown\cite{PhysLA_Lawal,CompConMat_Tse}.
%Physics Letters A 381 (2017) 2993–2999
%Computational Condensed Matter 4 (2015) 59e 63

In the next step we consider \VacSe{} native defects with concentration $x=0.075$ per formula unit. These defects act as electron donors, shifting the Fermi level upwards. Furthermore, it introduces mid-gap states inside the whole gap (Fig.~\ref{Fig:DOS_pure}(b)).
That behavior is different from the most common defects in \BiTe{} systems, Bi and Te antisites.

\subsubsection{ Mn dopants in the substitutional position (\MnBi{}) }

Our VASP based simulations reveal lattice relaxation around Mn dopants similarly to the case of \BiTe\cite{Carva2016_Mn-BiTe_FPTransport}. Within LDA and GGA-PBE approaches we obtained for 4$\%$ of \MnBi{}  the following relaxation of inter-atomic distances between Mn atoms and Bi, Se atoms. The Mn-Se distance shrinks from the initial 2.97 \AA{} to 2.60 \AA{} and the Mn-Bi distance 4.14 \AA{} is unchanged.  
In the TB-LMTO method the optimal choice to incorporate this relaxation was to assign the specie-dependent Wigner-Seitz radius $w^{\mathrm{Mn}}$ the Mn bulk native value (2.80 a.u.) instead of the Bi sublattice value, see also the Supplementary information of Ref.~\citenum{Carva2016_Mn-BiTe_FPTransport} for more details.

\MnBi{} acceptors shift the Fermi level to the valence band, where it crosses a small majority spin Mn impurity peak, acting as a virtual bound state. The size of this peak and its position w.r.t. valence band edge is highly sensitive to the amount of relaxation, as shown in Fig.~\ref{Fig:DOS_sub}(a) for sevaral different radii $w^{\mathrm{Mn}}$. The local \MnBi{} DOS is overall significantly modified by relaxation.  Only very little minority states are present at the Fermi energy. The band gap survives with its size reduced, contrary to the case of Mn-doped \BiTe{} \cite{Carva2016_Mn-BiTe_FPTransport}. The local \MnBi{} DOS contains also a minority peak slightly above the former conduction band edge, in agreement with the calculation based on the Korringa-Kohn-Rostoker Green function method
and the Dyson equation~\cite{Bouaziz2018_ImpurInducedStates}. We find that the unrelaxed DOS differs from the relaxed one, as well as the magnetic moment and other properties. 

The presence of \VacSe{} again induces mid-gap states. \EF{} is moved upwards and, interestingly, it can be fixed on the shallow mid-gap states inside the former gap. A low conductivity can be expected in this situation.

\subsubsection{ Mn dopants in the interstitial position (\Mni{}) }

Only small lattice relaxations were observed in the supercell based VASP calculations for \Mni{} atoms placed in the VdW gap, therefore no special treatment within LMTO approach was used for this case. States of interstitial Mn atoms fill the minority band gap. \Mni{} behaves as electron donor, oppositely to the effect of \MnBi{}. \EF{} now crosses the minority Mn electron peak, strongly hybridized with Se and Bi atoms, as shown in Fig.~\ref{Fig:DOS_int}.
For majority states a gap remains open, and only little electron states are present at the shifted \EF{}. This is basically independent of the \Mni{} concentration in the studied concentration range (1-10$\%)$. \Mni{} concentration increase leads only to an enhancement of virtual bound states.

%It is localized slightly above the majority gap, thus the material is not half-metal (?). We can assume obtaining half-metallic state by setting an appropriate concentration of \Mni{} , which move the \EF{} in the gap in the majority states. (?)

\VacSe{} defects do not cause a significant change of \EF{} position. The induced mid-gap states again remove the remaining gap in the majority channel.

%%%%%%%%%%%%%%%%%%%%%%%%%%%%%%%%%%%%%%%%%%%%%%%%%%%%%%%%%%%%%%%%%%%%%%%%%%%%%%%%%%%%%%%%%%%%%%%%%%%%%%%%%%%%5

\subsection{Magnetic moments}

%%%%%%%%%%%%%%%%%%%%%%%%%%%%%%%%%%%%%%%%%%%%%%%%%%%%%%%%%%%%%%%%%%%%%%%%%%%%%%%%%%%%%%%%%%%%%%%%%%%%%%%%%%%%%%%%%%%%%%%%%%%%%%%
%% package "booktabs" is NOT working there, although it is load at the beginning, it should ensure horizontal separation
\begin{table}[th!]
\begin{ruledtabular}
\begin{tabular}{cccccc}
\MnBi{}  & \Mni{} & \DefBiSe  & \VacSe  & $\mu$(\MnBi{})  & $\mu$(\Mni{}) \tabularnewline
\multicolumn{4}{c}{ conc. ($\times 10^{\mathrm{-2}}$ per f.u.)} & ($\mu_{B}$)  & ($\mu_{B}$) \tabularnewline
\hline
5  & - & - & -  & 4.03  & \tabularnewline
\;\,5$^{\ast}$  & - & - & -  & 4.59  & \tabularnewline
- & 5 & - & -  &  & 3.16 \tabularnewline
5  & - & 7.5 & -  & 3.91  & \tabularnewline
-  & 5  & 7.5 & - &  & 3.49 \tabularnewline
5  & - & - & 7.5  & 4.17  &  \tabularnewline
- & 5  & - & 7.5  &   & 3.24 \tabularnewline
2  & 4 & - & - & 4.17 & 3.25\tabularnewline
4  & 2 & - & - & 4.16 & 3.54\tabularnewline
\end{tabular}
\end{ruledtabular}

\caption{Selected values of calculated manganese magnetic moments $\mu$(\MnBi{})
and $\mu$(\Mni{}) for different concentrations of magnetic \MnBi
\, and \Mni \, together with a specified presence of native defects
(\DefBiSe, \VacSe). In the special case denoted by ({}$^{\ast}$) no relaxation around \MnBi{} was allowed.}
\label{tab:Mag_mom} 
\end{table}
%%%%%%%%%%%%%%%%%%%%%%%%%%%%%%%%%%%%%%%%%%%%%%%%%%%%%%%%%%%%%%%%%%%%%%%%%%%%%%%%%%%%%%%%%%%%%%%%%%%%%%%%%%%%%%%%%%%%%%%%%%%5555

%%%%%%%%%%%%%%%%%%%%%%%%%%%%%%%%%%%%%%%%%%%%%%%%%%%%%%%%%%%
\begin{figure*}
 \centering
 \includegraphics[width=1.95\columnwidth]{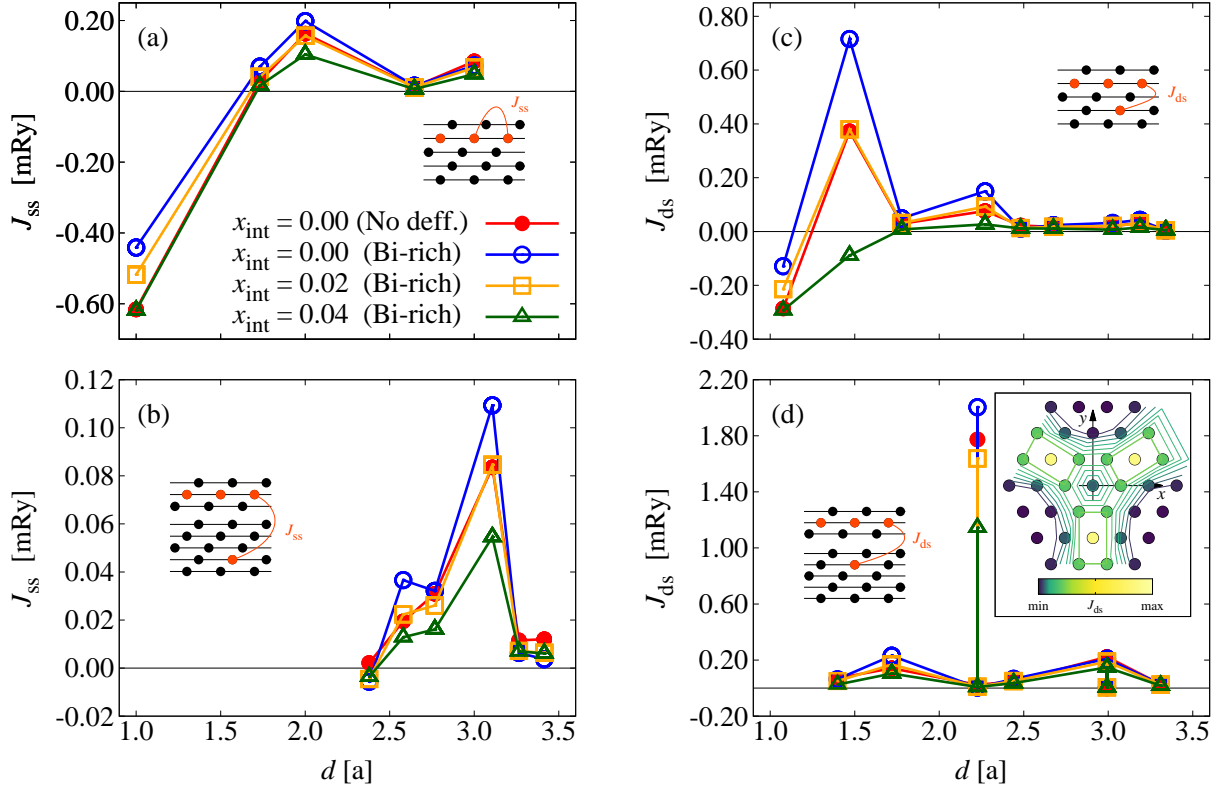}
 \caption{(Color online) Exchange interactions between magnetic moments of substitutional \MnBi{} atoms as a function of their distance $d$ (in the units of the lattice constant $a$): 
          (a) moments belonging to the same \MnBi{} sublattice of the same QL,
          (b) moments belonging to the same \MnBi{} sublattice in neighboring QLs,
          (c) moments belonging to different \MnBi{} sublattices of the same QL,
          (d) moments belonging to different \MnBi{} sublattices in neighboring QLs.
              Example positions of atoms taking place in that layer-wise subset of interactions are schematically depicted in each subfigure. The extra inset in (d) shows possible magnetic sites of one lattice layer, and its color corresponds to the intensity of its interaction with the atom in the next QL (its in-plane position is in the center), for the case $x_{\mathrm{int}}=0.00$. 
          The solid dots mark the interactions in case of no native defects and no interstitial Mn atoms ($\xint = 0.00$).
          The open symbols mark interactions in the presence of \DefBiSe{} native defects (Bi-rich form) for
          different concentrations of interstitial \Mni{} atoms: $\xint = 0.00$ (dots), 
          $\xint = 0.02$ (squares), $\xint = 0.04$ (triangles).}
 \label{Fig:exch1}
\end{figure*}
%%%%%%%%%%%%%%%%%%%%%%%%%%%%%%%%%%%%%%%%%%%%%%%%%%%%%%%%%%%

The values of magnetic moments for the \MnBi{} and \Mni{} atoms obtained from first principle calculations are significantly different (Tab. \ref{tab:Mag_mom}) and depend rather weakly on concentrations of Mn atoms and native defects studied in this paper.
The magnetic moment of \MnBi{} atoms slowly shrinks with the increasing concentration of \MnBi{} atoms, from the zero concentration limit value of $4.18~\muB$ (where $\muB$ denotes the Bohr magneton). %where at the concentration of \MnBi{} $x=0.10 / f.u.$ obtains the value of $4.96\, \muB$.
 This magnetic moment slightly increases with an addition of \Mni{} or \VacSe{}. An inverse effect was observed in the presence of \DefBiSe{}, which suppresses \MnBi{} magnetic moments, especially for low \MnBi{} concentration. Similarly to what we have found for DOSes, Mn magnetic moment would be significantly modified if the relaxation was not taken into account.

Magnetic moment magnitudes of \Mni{} atoms fit in the range 3.21 - 3.15~$\muB$ for \Mni{} concentrations between 1\% and 10\%. All the investigated native defects enhance magnetic moments of \Mni{} atoms, this effect is most pronounced for the presence of \DefBiSe, where the obtained moment values are close to $3.50~\muB$, depending on the concentration.

The low variation of moment magnitude enables us to approximate the value of magnetic moment by a fixed value, achieving a deviation less than two percent from the calculated value in the considered region of the concentrations.
Thus in our spin dynamics simulations we have always used magnetic moments $4.00~\muB$ for \MnBi{} and $3.50~\muB$ for \Mni{} atoms.

\subsection{Exchange interactions}

%%%%%%%%%%%%%%%%%%%%%%%%%%%%%%%%%%%%%%%%%%%%%%%%%%%%%%%%%%%
\begin{figure*}[ht!]
 \centering
 \includegraphics[width=1.95\columnwidth]{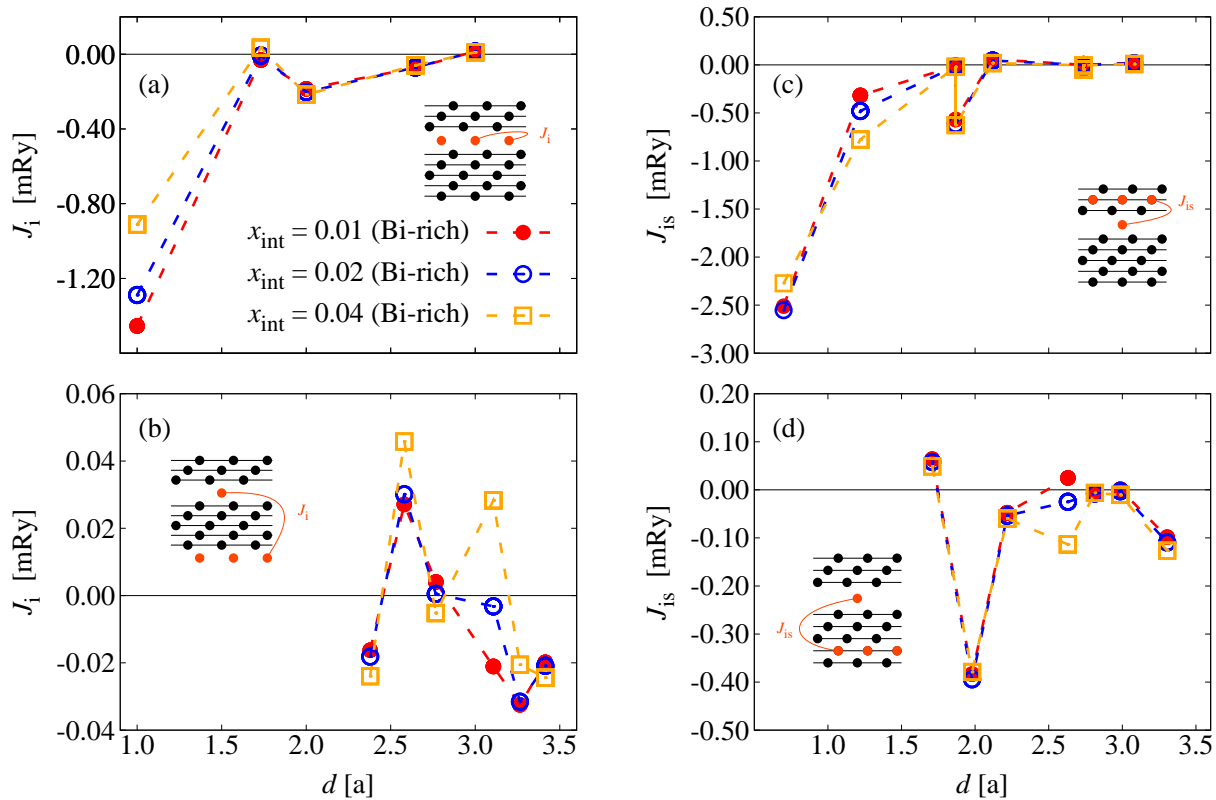}
 \caption{(Color online) Exchange interactions between magnetic moments of interstitial \Mni{} atoms for:
          (a) \Mni{} atoms in the same vdW gap, (b) \Mni{} atoms in neighboring vdW gaps.
          Exchange interactions between magnetic moments interstitial \Mni{} and substitutional \MnBi{} atoms for:
          (c) \MnBi{} atoms on the adjacent side of the QL, (d) \MnBi{} atoms on the opposite side of the QL.
          The symbols mark different concentration of interstitial \Mni{} atoms:
          $\xint = 0.01$ (solid dots), $\xint = 0.02$ (open dots), and $\xint = 0.04$ (open squares).}
 \label{Fig:exch2}
\end{figure*}
%%%%%%%%%%%%%%%%%%%%%%%%%%%%%%%%%%%%%%%%%%%%%%%%%%%%%%%%%%%

Calculated exchange interactions between \MnBi{} atoms are shown in Fig.~\ref{Fig:exch1} as a function of their distance $|\Dr|$. These are split into interactions between Mn moments on the same (a,b) and different (c,d) Bi sublattices, and also into interactions between Mn located inside one QL (a,c) and at different QLs (b,d). The interactions are plotted for Mn atoms in \BiSe{} without any native defects,
as well as \BiSe{} with \DefBiSe{}. Moreover, for the Bi-rich form of \BiSe{} we assumed different concentrations of \Mni{} atoms in
the vdW gap. 

First of all let us note that in the same Mn sublattice (Fig.~\ref{Fig:exch1}(a)) the strongest interaction between the 
closest possible Mn atoms is always negative (antiferromagnetic). This interaction also varies with the presence of native defects and magnetic impurities.
Although this interaction is the strongest one, in dilute systems one generally needs to consider a larger amount of neighbors and the importance of the interaction to the nearest neighbor is limited due to the magnetic percolation effect \cite{r_10_Sato_Kud_Turek_FP_DMS}.
The other values of $\Jss$ are positive. The other positive interactions may overcome that of the negative first interaction, and we show that this indeed happens for selected cases.
In the case of no interstitial Mn atoms, the presence of \DefBiSe{} defects shifts the values of $\Jss$ in the positive direction.
In turn the effect of \Mni{} atoms seems to be opposite. With increasing $\xint$, the values of $\Jss$ are lowered.
Similar trends can be observed for the values of $\Jss$ acting between different QLs, shown on Fig.~\ref{Fig:exch1}(b).
There is only one negative interaction, which is, however, very small. The strongest interactions are positive.

Let us now concentrate on the interactions between substitutional Mn atoms belonging to different magnetic sublattices.
Similarly to $\Jss$, also the first value of $\Jds$ is negative (Fig.~\ref{Fig:exch1}(c)). 
In most of the studied cases, the strongest $\Jds$ interaction inside QL is the second one, which is positive just in the case of small concentrations of \MnBi{} atoms ($\xint = 0$ or $0.02$).
Generally, $\Jds$ values inside QL decay relatively quickly with distance, with values approaching zero already for distances above 2.5.
When the interstitial \Mni{} atom concentration exceeds certain critical value ($\xint = 0.04$), all the significant $\Jds$ interactions turn to be negative and decrease almost exponentially towards zero with only little deviations.

%%%%%%%%%%%%%%%%%%%%%%%%%%%%%%%%%%%%%%%%%%%%%%%%%%%%%%%%%%%
\begin{figure*}[ht!]
 \centering
 \includegraphics[width=0.7\textwidth]{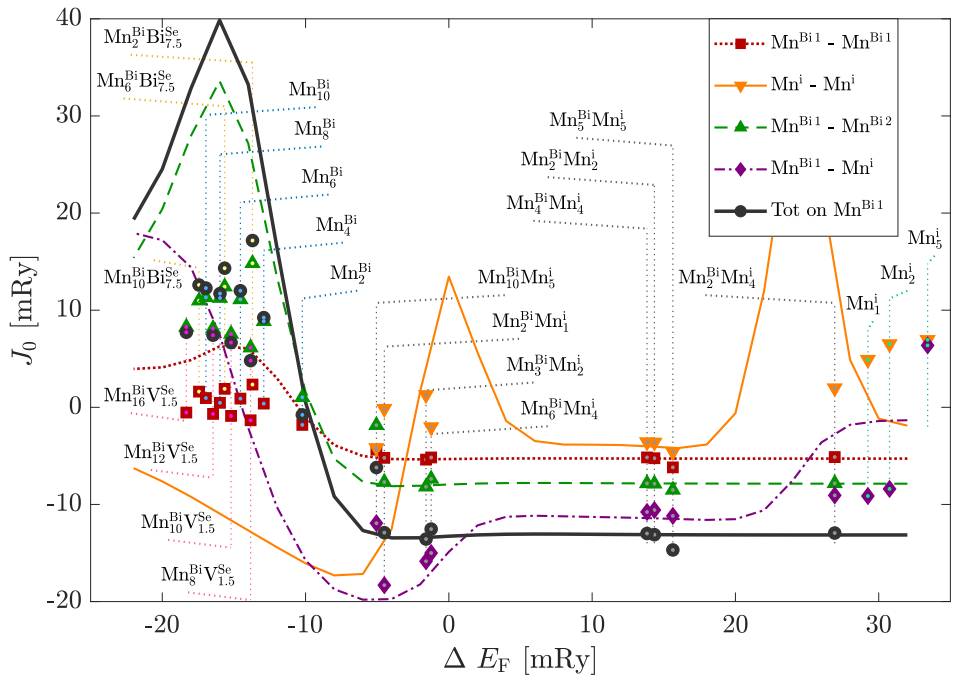}
 \caption{(Color online) Lines: Exchange interactions between different magnetic dopants as a function of the Fermi level position relative to the Fermi level of pure \BiSe{}, summed over all considered shells. Points: exchange interactions between different magnetic dopants calculated for a specific combination of dopants (see text), depicted at the energy corresponding to its own Fermi level (summed over all considered shells). Subscripts of dopants' labels specify dopant concentration in percents per f.u.} 
 \label{Fig:exch_sums}
\end{figure*}
%%%%%%%%%%%%%%%%%%%%%%%%%%%%%%%%%%%%%%%%%%%%%%%%%%%%%%%%%%%

Most interesting results concern $\Jds$ interactions acting between Mn atoms located in different QLs (Fig.~\ref{Fig:exch1}(d)).
Almost all of these interactions are positive. The by far largest value of $\Jds$ is obtained for the distance  $d \simeq 2.23\, a$, and it is actually the largest from all the calculated interaction values. In order to understand this anomaly, positions of possible magnetic sites (Bi sublattice) of one lattice layer are depicted in the inset of Fig.~\ref{Fig:exch1}(d) together with the $\Jds$ interaction intensity. Depicted exchange interactions fulfill the three-fold symmetry as expected. \MnBi{}-\MnBi{} connections with the maximum interaction value actually pass very close to Se atoms located in the layers between them. Along the \MnBi{}-Se-Se-\MnBi{} chain, both \MnBi{}-Se distances are only 0.7$a$, and Se-Se distance is 0.85$a$ (across the vdW gap). Se presence it thus sufficient to enhance the interaction even above the value corresponding to the closest \MnBi{}-\MnBi{} distance $a$. 
For the unrelaxed \MnBi{} we obtain significantly different exchange interactions than for the relaxed ones. This is probably the root cause for the difference from the results of a previous calculation \cite{r_14_Vergn_Mertig_ExchInt_BiSe_BiTe_SbTe}, where relaxations were neglected.

Fig.~\ref{Fig:exch2} describes the exchange coupling involving \Mni{} atoms located in the vdW gaps.
Figs.~\ref{Fig:exch2}(a) and (b) plot the interactions between \Mni{} type magnetic moments. Inside one vdW gap one can
notice a dominance of negative $\Ji$ values. 
The largest interaction, which is most sensitive to concentration of \Mni{} atoms, is the nearest neighbor contribution. 
The absolute value of this interaction decreases with increasing $\xint$.
The coupling strength between \Mni{} atoms belonging to neighboring vdW gaps, shown in Fig.~\ref{Fig:exch2}(b), 
features both positive and negative values of $\Ji$.
The interactions between magnetic moments of \Mni{} and \MnBi{}, $\Jis$, atoms are plotted on Figs.~\ref{Fig:exch2}(c) and (d). 
The overwhelming majority of them are again negative. Especially noticeable are the relatively strong coupling
strengths between close \Mni{} and \MnBi{} moments.

Already from these figures one can see that the dependence of exchange interactions on the presence of various defects/dopants is strong and complex. Considering the vast number of possible defect combinations here it would appear as difficult to map the possible behavior of exchange interactions in this configuration space. However, a significant  contribution to exchange could be due to conduction electrons, which would depend mostly on the position of the Fermi level. Therefore we have decided to examine exchange interactions as a function of the Fermi level, described by its shift $\delta E$ from its value for the unperturbed system. The corresponding calculations of exchange interactions for a number of possible Fermi level values were performed for the reference system with just two magnetic dopants in the unperturbed \BiSe{}, which formally corresponds to the limit of zero concentration of magnetic dopants treated in the framework of the CPA. These can also be calculated from the Lichtenstein formula (Eq.~\ref{Eq:Liecht}). We show here only sums of exchange interaction over all the considered neighbors, $J^{\Dop\Dop'}_{0}=\sum_{\Dr} J^{\Dop\Dop'}(\Dr)$, instead of interactions for specific distances. This is the most relevant quantity in a dilute medium, and its use makes the whole comparison feasible. In Fig.~\ref{Fig:exch_sums} the sums are plotted for the reference system with zero defect concentration and just a shifted Fermi level, together with sums calculated for systems with specific combinations of defects. The Fermi level shift on the $x$ axis represents an input parameter for the first type of calculation, while for the latter it is determined from the self-consistent cycle.  The agreement appears to be rather good, especially the shapes of dependencies are similar for both types of calculations. Few smaller discrepancies arise, mainly overall weaker exchange interactions were obtained for truly disordered systems than for the reference system. This reduction can be due to increased electron scattering. The correspondence with Fermi level shifted systems indicates the importance of RKKY interaction relying on conduction electrons. 

We can now obtain some insight into the overall evolution of magnetic interactions with doping. Most important here is the increase of interactions between \MnBi{} atoms for $p$-doped samples with $E_F$ in the range of Fermi level shift $\delta E<-10~\mathrm{mRy}$, where positive exchange interactions appear, suggesting a possibility of a more stable FM order there. This situation is expected in the case of pure \MnBi{} doping and also \DefBiSe{} antisites.
The tendency to order might however be reduced by the presence of \Mni{} and \VacSe{} dopants, which both shift the Fermi level upwards. Furthermore, \Mni{} moments provide a strong negative coupling to \MnBi{} moments.

%%%%%%%%%%%%%%%%%%%%%%%%%%%%%%%%%%%%%%%%%%%%%%%%%%%%%%%%%%%
\begin{figure}[ht!]
  \centering
  \includegraphics[width=.9\columnwidth]{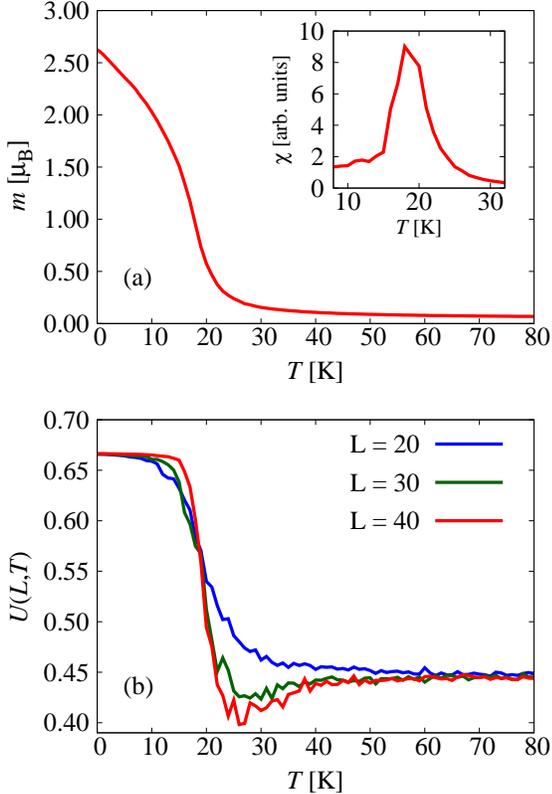}
  % original
  %\caption{(Color online) Example of temperature dependence of quantities calculated using atomistic MC simulations
  %          for a Bi-rich Mn-doped \BiSe{} with $\xint = 0.00$ and $\xsub = 0.10$: (a) magnetization, 
  %          and magnetic susceptibility (in arb. units) in the inset, (b) Binder cumulants for different sizes of simulated supercell:
  %          $L = 20$, $30$, and $40$.}
  % changed           
  \caption{(Color online) Example of temperature dependence of quantities calculated using atomistic MC simulations
            for a Bi-rich Mn-doped \BiSe{} with $\xint = 0.00$ and $\xsub = 0.10$: (a) magnetization $m$, 
            and magnetic susceptibility $\chi$ in the inset, (b) Binder cumulants $U$ for different sizes of simulated supercell:
            $L = 20$, $30$, and $40$.}         
  \label{Fig:MagDyn}

\end{figure}
%%%%%%%%%%%%%%%%%%%%%%%%%%%%%%%%%%%%%%%%%%%%%%%%%%%%%%%%%%%

\subsection{Curie temperatures}

%%%%%%%%%%%%%%%%%%%%%%%%%%%%%%%%%%%%%%%%%%%%%%%%%%%%%%%%%%%
\begin{figure}[ht!]
  \centering
  \includegraphics[width=0.99\columnwidth]{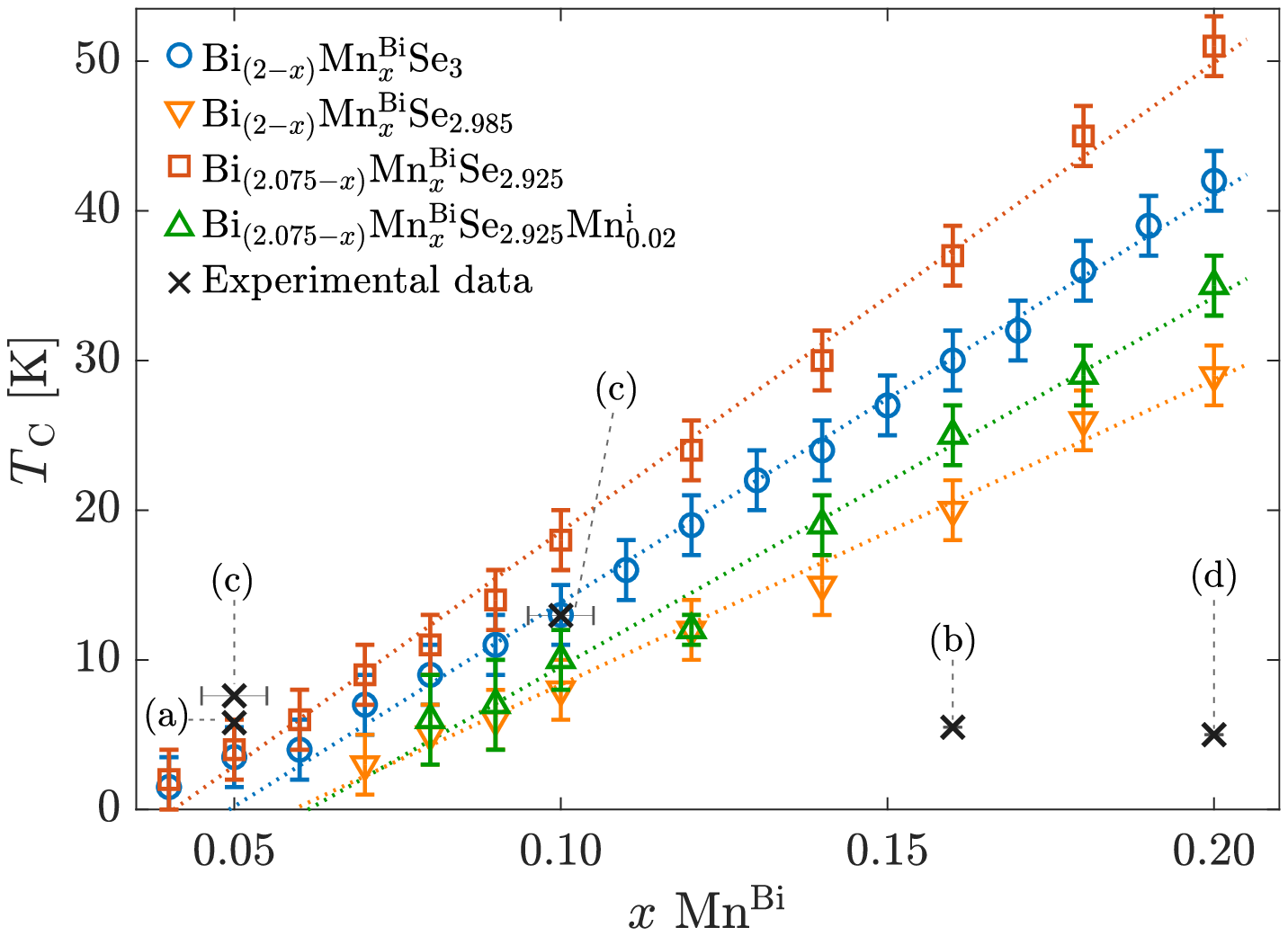}
  \caption{(Color online) Dependence of the Curie temperature on the concentration $\xsub$ of substitutional \MnBi{} atoms, and fits to the calculated points. Depicted data: ({$\bigcirc$}) \MnBi{} defects only, ($\Box$) \MnBi{} and \DefBiSe  defects, ($\triangle$) previous case together with \Mni{}, ($\nabla$) \MnBi{} defects with \VacSe, ($\times$) experimental results denoted by letters (a)-(c) originate from three references as follows: a\cite{r_13_Barde_FM_Mn-BiSe_epitaxial}, b\cite{r_16_Sanch_Springholz_Holy_NonmBandGap_Mn-BiSe}, c\cite{r18_Maurya_JMMM456}.}
  %\caption{(Color online) Dependence of the Curie temperature on concentration of the substitutional \MnBi{} atoms per f.u. $\xsub$  and fits to the calculated points. Depicted data: (blue) \MnBi{} defects only, (red) \MnBi{} and \DefBiSe  defects, (green) previous case together with \Mni{}, (orange) \MnBi{} defects with \VacSe, (gray) experimental data: (multilayers) a\cite{r_13_Barde_FM_Mn-BiSe_epitaxial},b\cite{r_16_Sanch_Springholz_Holy_NonmBandGap_Mn-BiSe} (bulk single crystal) c\cite{r18_Maurya_JMMM456} \textbf{stara verze - exp. data obtizne pouzitelna: a\cite{r_16_}Sanch_Springholz_Holy_NonmBandGap_Mn-BiSe}, b\cite{r_13_Barde_FM_Mn-BiSe_epitaxial},   c\cite{Duming_Zhang:PRB_2012}, d\cite{r_16_Taras_cb_MagStruct_Mn_BiSe}, e\cite{AIPAdv14_Collins}, f\cite{r_16_Valis_n-p_Mn-BiSe_Hetero}, (bulk single crystal) g\cite{r18_Maurya_JMMM456}}.}
  %~\cite{Duming_Zhang:PRB_2012,r_13_Barde_FM_Mn-BiSe_epitaxial,r_16_Sanch_Springholz_Holy_NonmBandGap_Mn-BiSe,AIPAdv14_Collins}}
  \label{Fig:Tc}
\end{figure}
%%%%%%%%%%%%%%%%%%%%%%%%%%%%%%%%%%%%%%%%%%%%%%%%%%%%%%%%%%%
Let us now study the finite temperature magnetism in Mn-doped \BiSe{} using MC simulations with exchange coupling strengths 
introduced in the previous subsection.
Fig.~\ref{Fig:MagDyn}(a) shows an example of averaged physical quantities calculated using atomistic MC simulations as a function of temperature
for a Mn-doped Bi-rich \BiSe{} including just substitutional Mn atoms ($\xint = 0.00$). In the presented case, we assumed concentration of \MnBi{} atoms $\xsub = 0.10$.
Magnetization curve has a shape typical for a ferromagnetic phase transition with Curie temperature $\Tc \simeq 18~{\rm K}$. 
This observation is supported by magnetic susceptibility showing a sharp peak at this temperature. Finally, Fig.~\ref{Fig:MagDyn}(b) shows Binder cumulants
calculated for three different sizes of the supercell, $L = 20$, $30$, and $40$. All the Binder cumulants intersect at the same temperature, which
allow us to estimate the Curie temperature to be $\Tc \simeq 18~{\rm K}$.
The same method can be used to study $\Tc$ for various concentrations of \Mni{} and \MnBi{} atoms as well as native defects.

Fig.~\ref{Fig:Tc} shows the calculated Curie temperature as a function of substitutional \MnBi{} atoms concentrations, $\xsub$.
At first sight, one can notice the linear dependence of $\Tc$ on $\xsub$ for all the studied cases.
%The dashed lines present to linear fits to the calculated points.
%The solid dots mark the case without any native defects. At small concentrations of \MnBi{} atoms the critical temperature approaches to $\Tc \simeq 10\, {\rm K}$. 
In simulations it was impossible to obtain magnetic ordering for concentrations of \MnBi{} atoms less than $x=0.04$. It is caused not only by the decreasing intensity of exchange interactions $J_{ij}$ (Fig.~\ref{Fig:exch_sums}), but probably also by the vicinity of the percolation limit.  
% Nejsme v tomto rezimu: Notably, Mn concentrations above $x=0.3$ (6 atomic \%) lead to significant distortions of the lattice \cite{r_16_Taras_cb_MagStruct_Mn_BiSe} and this would have to be taken into account when calculated in such a regime.

Inclusion of \DefBiSe{} native defects causes a remarkable increase of $\Tc$  (Fig.~\ref{Fig:Tc}). 
The temperatures are shifted to higher values and the slope of the linear dependence becomes steeper.
Further increase of \Mni{} concentration reduces the critical temperature as well as the slope, similarly to the presence of \VacSe{} defects. This behavior correlates with our findings based on studying the interaction sums $J_0$. The few existing experimentally measured values of $\Tc$ (shown in Fig.~\ref{Fig:Tc}) appear to agree with our predictions if the presence of a sufficient amount of undetermined defects (\Mni{} and \VacSe{}) is assumed. Notably, in one case ferromagnetic order has been observed only after doping has changed the sample conductivity from n-type to p-type \cite{r18_Maurya_JMMM456}, in agreement with our predictions about Fermi level dependence.

\section{Conclusions}
We have calculated the electronic structure and exchange interactions in Mn-doped \BiSe{} employing first-principles methods. Apart from the mostly expected Mn position substituting Bi atoms we also considered interstitial Mn, and find that in this position it has strikingly different effect. 
Our calculations predict a significant relaxation of atoms surrounding substitutional Mn. Magnetism-related properties are sensitive to it, and its neglect would lead to a significantly different prediction of exchange interactions or magnetic moment. 

Interestingly, although the nearest neighbor interaction is always negative in the studied systems, in some cases interactions to more distant neighbors overcome this contribution and the system then prefers ferromagnetic ordering.
We have found that in the studied cases the strongest interaction corresponds to a specific pair of Mn atoms with the distance 2.23 higher than that of nearest neighbors. This quite unusual behavior is a consequence of the lattice structure and cannot be expected for example in the commonly studied cubic (Ga,Mn)As dilute magnetic semiconductors. This also means that the interaction range cannot be limited to just few nearest neighbors here, we recommend to include shells up to the distance 3$a$. For this reason we also do not expect any 2D magnetic arrangement along layers, as already noticed in a recent experiment \cite{Savchenko2018_MomentForm_Mn-doped-BiSe_FMR}.

Magnetic ordering as a function of temperature was investigated. Our spin dynamics simulations predict the existence of a phase transition to FM order for a wide range of Mn-doping concentrations. In the interesting concentration range $x$ = 0.05 - 0.1 a linear increase of $\Tc$ with the concentration of substitutional Mn is obtained, in agreement with the preceding works \cite{r_14_Vergn_Mertig_ExchInt_BiSe_BiTe_SbTe}. Furthermore, we find that an increase of interstitial Mn has the opposite effect and decreases the Curie temperature. A plausible scenario for the experimentally observed weak dependence of $\Tc$ on Mn concentration could thus be based on a simultaneous increase of both \MnBi{} and \Mni{}.
Another explanation may be due to the influence of Mn doping on native defects, which has already been suggested \cite{r_08_Janic_JVej_Secho_Transp_Magn_Mn-BiSe_single, r_15_Babak_Taraf_StructProp_MnDop_BiSe_MBE, r_16_Wolos_HS_Mn_Bi2Se3_Vac}. Therefore we also performed calculations of the system in the presence of the most probable native impurities. We find that the presence of Se vacancies decreases $\Tc$, while \DefBiSe{} antisites increase it. A sufficient concentration of these extra defects can also prevent formation of FM order at all. The order of magnitude of variations of Tc for constant Mn concentrations is consistent with the experimentally reported ones and underline of the importance of coexistence of different defect types.

The total variation of exchange interactions and the resulting evolution of $\Tc$ when varying dopant concentrations can be largely explained as the consequence of the shift of Fermi level in this system. For small Fermi level shifts due to dopants the sums over exchange parameters show an overall increase with decreasing Fermi level, hence we expect a more stable FM order and a higher $\Tc$ for p-type samples. Notably, the most common defects here, Se vacancies, shift the Fermi level to higher energies. Therefore counterdoping is needed to obtain p-type samples, and its proper application can lead to an increase of $\Tc$.

%%%%%%%%%%%%%%%%%%%%%%%%%%%%%%%%%%%%%%%%%%%%%%%%%%%%%%%%%%%
%%%%%%%%%%%%%%%%%%%%%%%%%%%%%%%%%%%%%%%%%%%%%%%%%%%%%%%%%%%

\label{Sec:Conclusions}

\begin{acknowledgments}
  This work was supported by the Czech Science Foundation Grant no. (19-13659S). 
  %%% MetaCentrum
  Access to computing and storage facilities owned by parties and projects contributing to the National Grid Infrastructure MetaCentrum, 
  provided under the program "Projects of Large Research, Development, and Innovations Infrastructures" (CESNET LM2015042), is greatly appreciated.
\end{acknowledgments}

%\clearpage
\bibliographystyle{apsrev}
\bibliography{b_kc,b_pb,b_js}

%\printbibliography

\end{document}